Manuscript (including abstract)

# Partial Residuated Implications Derived from Partial Triangular Norms and Partial Residuated Lattices*

Xiaohong Zhang [a, b ♦], Nan Sheng [a], Rajab Ali Borzooei [c]

[a] Department of Mathematics, Shaanxi University of Science & Technology, Xi'an 710021, China

[b] Shaanxi Joint Laboratory of Artificial Intelligence, Shaanxi University of Science & Technology, Xi'an 710021, China

[c] Department of Mathematics, Faculty at Mathematical Sciences, Shahid Beheshti University, Tehran 1983963113, Iran

**Abstract**

In this paper, we reveal some relations between fuzzy logic and quantum logic, and mainly study the partial residuated implications (PRIs) derived from partial triangular norms (partial t-norms) and partial residuated lattices (PRLs), and expand some results in the article "material implication in lattice effect algebra". Firstly, according to the concept of partial triangular norms given by Borzooei, we introduce the connection between lattice effect algebra and partial t-norms, and prove that partial operations in any commutative quasiresiduated lattice are partial t-norms. Secondly, we give the general form of partial residuated implications and the concept of partial fuzzy implications (PFIs), and the condition that partial residuated implication is a fuzzy implication is given. We also prove that each partial residuated implication is a partial fuzzy implication. Thirdly, we propose the partial residuated lattice and study their basic properties, to discuss the corresponding relationship between PRLs and lattice effect algebras (LEAs), to further reveal the relationship between LEAs and residuated partial algebras. In addition, like the definition of partial t-norms, we also propose the concepts of partial triangular conorms (partial t-conorms) and corresponding partial co-residuated lattices (PcRLs). Finally, based on partial residuated lattices, we give the definition of well partial residuated lattices (wPRLs), study the filter of well partial residuated lattices, and then construct quotient structure of partial residuated lattices.

## 1. Introduction

In 1965, Zadeh first put forward the term fuzzy set in [22], marking the birth of fuzzy mathematics. Then, he gradually established fuzzy logic. In fuzzy logic, the importance of t-norms and t-conorms is self-evident. Therefore, scholars have an endless stream of research on them. We know that the operation involved in t-norms is a conventional binary operation, and so is t-conorms. However, in the practical application of fuzzy logic, there will be some "undefined" situations. In other words, in many cases, it does not involve propositions that are true to some extent, and it may not involve propositions that are true to any degree extent. In fact, from the point of view of partial membership function and fuzzy partial logic (logical connectives are partial operations), some

* This work was supported by National Natural Science Foundation of China (Nos. 61976130, 62081240416).

♦ Corresponding author (email: zhangxiaohong@sust.edu.cn or zxhonghz@263.net)

scholars have explored and studied in [2,3,17,18], for example, in [2], Běhounek and Novák discuss fuzzy partial logics, in which they use the special value "*" to explain semantics such as "undefined", "meaningless", "non-applicable", etc. They give the definitions of unary and binary original conjunctions in the following tables:

| c | $\beta$ | * |
|---|---|---|
| $\alpha$ | $\alpha c \beta$ | 0 |
| * | * | * |

| $\wedge$ | 0 | $\delta$ | * |
|---|---|---|---|
| 0 | 0 | 0 | 0 |
| $\gamma$ | 0 | $\gamma \wedge \delta$ | * |
| * | 0 | * | * |

In [4,6], Burmeister and Běhounek also provide methods to deal with "undefined" cases from the perspective of aggregate functions and partial algebras, respectively.

In 1994, Foulis and others put forward the effect algebras that can describe imprecise quantum phenomena in [12]. Among them, the binary operation "⊕" in the effect algebra they defined is also a partial operation. Therefore, whether in fuzzy logic or quantum logic, partial operation is an important research direction, which can express a wider range of uncertainty. Then, for t-norms, Borzooei and Dvurečenskij and others proposed the concept of partial t-norm when studying lattice valued quantum effect algebra in [5] in 2018. The key is to replace the conventional binary operation in t-norm with "partial binary operation", that is, the operation between some "element pairs" is allowed to be "undefined". In [15], Wei gave the concept of partial t-implication through partial t-norm. But, none of them has further studied partial t-norms. In [8], Chajda and Halas and others investigate the natural implications in lattice effect algebras, and effect implication algebras were also studied by them. In recent years, the study of fuzzy implications and residuated implications in effect algebra is also favored by scholars (see [5,9,15,21,23]). All these provide important ideas for us to study partial t-norms. Therefore, we can also similarly study partial t-norms and their residuated (partial) implications, including their algebraic characterization. For t-conorms, in 1997, Baets puts forward the concept of coimplication in [1]. In [24], Zheng and Wang introduced structure (⊕,⊖) on lattice semigroups, and established the theory of co-residuated lattices. However, partial algebraic structures of t-conorms have not been excavated. Regarding the algebraic structure of partial t-norms, it must be mentioned that in [25], Zhou and Li proposed the concept of PRL (we call it $z\mathcal{L}$-PRL). With the in-depth study of fuzzy logic and quantum logic, the relationship between them has been widely concerned (see [13,20,26,27,28]). Zhou and Li also proved that PRLs can be considered as the common generalization of residuated lattice structure in fuzzy logic and a lattice effect algebra in quantum logic. However, we strictly prove that according to the definition of this paper, partial operation must be full operation. Therefore, in order to properly characterize partial t-norms and their residuated implications, we will redefine partial residuated lattices and reasonably study the relationship between them and lattice effect algebras. In addition, the filter and congruence relations between residuated lattice and effect algebra are studied (see [7,10,11,16,19]). Then, we can learn from them and study the relevant contents of PRLs.

The paper will be carried out from the following aspects: First, we give the general form of partial residuated implications derived from partial t-norms, and reasonably define partial fuzzy implications. Second, the concept of partial adjoint pairs (PAPs) is properly defined. Based on this, partial residuated lattices are defined. We also establish the induction relationship between lattice effect algebra and PRL. Finally, the filter and quotient structure of PRLs are established.

## 2. Lattice effect algebras and partial t-norms

**Definition 2.1**[9,12,15] Assume $E = (E, +, ', 0, 1)$ is an effect algebra, with a partial binary operation $+$ and a unary operation $'$. If the following hold:

(E1) $x + y$ is defined iff $y + x$ is defined, and then $x + y = y + x$;

(E2) $x + y$ and $(x + y) + z$ is defined iff $y + z$ and $x + (y + z)$ are defined, and then $(x + y) + z = x + (y + z)$;

(E3) A unique $x' \in E$ with $x + x' = 1$;

(E4) if $x + 1$ is defined, then $x = 0$.

Define a partial order $\leq$ on $E$ by $x \leq y$ iff there exists an element $z \in E$ and $x + z = y$. For all $x \in E$, $0 \leq x \leq 1$. And define a partial operation – by $x \leq y$ and $y - x \coloneqq z$ iff $x + z$ is defined and $x + z = y$. We call $E$ is an LEA, if the effect algebra $E$ is also a lattice under $\leq$.

**Theorem 2.2**[9] If $E = (E, \leq, +, ', 0, 1)$ is an LEA. Then the following hold:

(1) $x + y$ is defined iff $x \leq y'$.

(2) If $x \leq y$ and $y + z$ is defined, then $x + z$ is defined and $x + z \leq y + z$.

(3) If $x \leq y$, then $x + (x + y')' = y$.

**Definition 2.3**[9] A partial algebra $\boldsymbol{C} = (C, \vee, \wedge, \odot, \rightarrow, 0, 1)$, where $(C, \vee, \wedge, 0, 1)$ is a bounded lattice and $\odot$ is a partial operation, $\rightarrow$ is a full operation, is called a commutative quasiresiduated lattice. If the following hold:

(i) $(C, \odot, 1)$ is a commutative partial monoid and $x \odot y$ is defined iff $x' \leq y$;

(ii) $x'' = x$ and if $x \leq y$, then $y' \leq x'$;

(iii) $(x \vee y') \odot y \leq y \wedge z$ iff $x \vee y' \leq y \rightarrow z$.

**Theorem 2.4**[23] Assume $(C, \vee, \wedge, \odot, \rightarrow, 0, 1)$ is a commutative quasiresiduated lattice. Then the following hold:

(1) $x' \leq y$ implies $x \odot y \leq y$.

(2) $x' \leq y$ implies $x \leq y \rightarrow (x \odot y)$.

(3) If $x' \leq y$ and $z \leq y$, then $x \odot y \leq z$ iff $x \leq y \rightarrow z$.

**Definition 2.5**[5] The binary operation $\odot$ is a partial operation on a bounded lattice $L$, it is a partial t-norm, if,

(1) $x \odot 1 = x$;

(2) If $x \odot y$ is defined, then $y \odot x$ is defined and $x \odot y = y \odot x$;

(3) If $y \odot z$ and $x \odot (y \odot z)$ are defined, then $x \odot y$ and $(x \odot y) \odot z$ are defined and $x \odot (y \odot z) = (x \odot y) \odot z$;

(4) If $x \leq y$, $h \leq k$ and $x \odot h$, $y \odot k$ are defined, then $x \odot h \leq y \odot k$.

**Example 2.6** Assume $L$ = [0,1]. Define ⊙ as follows:

$$a \odot b = \begin{cases} undefined, & if\ a, b \in [0,0.5] \\ min\{a, b\}, & others \end{cases}$$

Then the operation ⊙ is a partial t-norm.

**Example 2.7** Assume $L$= [0,1]. Define ⊙ as:

$$x \odot y = \begin{cases} min\{x, y\}, & if\ x, y \in [0.5,1]\ or\ x = 1\ or\ y = 1 \\ undefined, & others \end{cases}$$

Then the operation ⊙ is a partial t-norm.

**Example 2.8** Assume $L$= [0,1]. Define ⊙ as:

$$x \odot y = \begin{cases} x \wedge y, & if\ x + y \leq 1\ or\ x = 1\ or\ y = 1 \\ undefined, & others \end{cases}$$

Then the operation ⊙ is a partial t-norm.

**Example 2.9** Assume $L$= [0,1]. Define ⊙ as:

$$x \odot y = \begin{cases} x \wedge y, & if\ x + y \leq 0.5\ or\ x = 1\ or\ y = 1 \\ undefined, & others \end{cases}$$

Then the operation ⊙ is a partial t-norm.

**Example 2.10** Assume $L$= [0,1]. Define ⊙ as:

$$x \odot y = \begin{cases} x \wedge y, & if\ x + y \leq \alpha\ or\ x = 1\ or\ y = 1 \\ undefined, & others \end{cases}$$

where $\alpha \in [0,1]$. Then the operation ⊙ is a partial t-norm.

**Example 2.11** Assume $L$ = {0,1,2,3,4}, Fig. 1 shows the partial order ≤. Operation ⊙ is defined by Table 1. Then ⊙ is a partial t-norm.

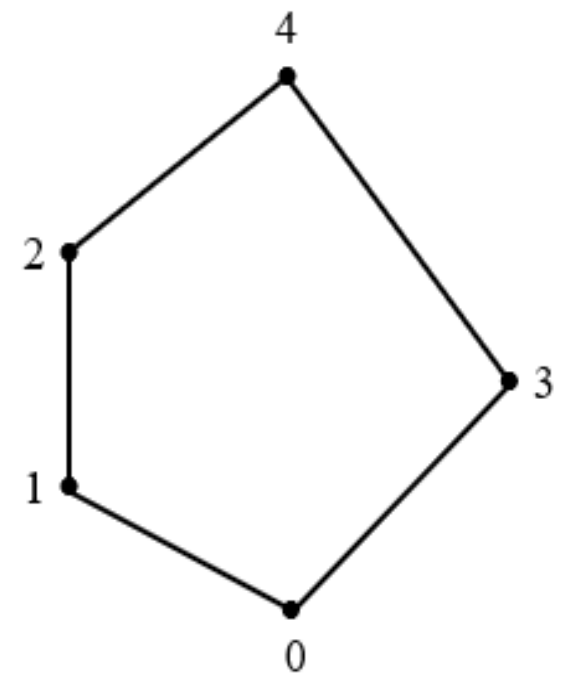

Fig. 1. Order relation

Table 1. Partial operation ⊙

| ⊙ | 0 | 1 | 2 | 3 | 4 |
|---|---|---|---|---|---|
| 0 | | | | | 0 |
| 1 | | | 0 | | 1 |
| 2 | | 0 | 1 | | 2 |
| 3 | | | | 0 | 3 |
| 4 | 0 | 1 | 2 | 3 | 4 |

**Example 2.12** Assume $L = \{0,1,2,3,4\}$, Fig. 2 shows the partial order $\leq$. Operation ⊙ is defined by Table 2. Then ⊙ is a partial t-norm.

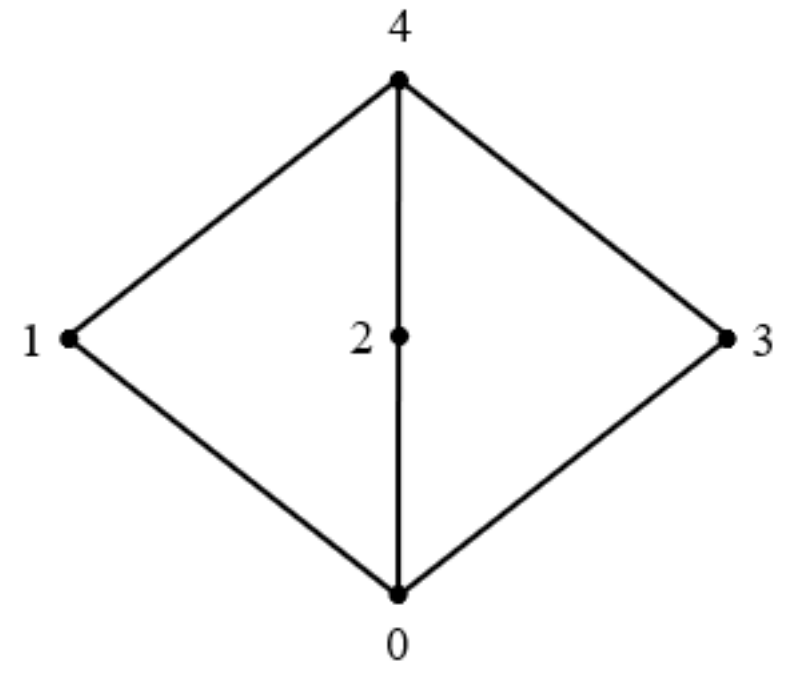

Fig. 2. Order relation

Table 2. Partial operation ⊙

| ⊙ | 0 | 1 | 2 | 3 | 4 |
|---|---|---|---|---|---|
| 0 | | | | | 0 |
| 1 | | | | 0 | 1 |
| 2 | | | 0 | | 2 |
| 3 | | 0 | | | 3 |
| 4 | 0 | 1 | 2 | 3 | 4 |

**Proposition 2.13** If $E = (E, \leq, +, ', 0, 1)$ is an LEA, suppose,

$$x \odot y := (x' + y')' \text{ iff } x' \leq y.$$

Then ⊙ is a partial t-norm.

**Proof.** (1) Since $x' \leq 1$, so $x \odot 1$ is defined, then $x \odot 1 = (x' + 1')' = (x' + 0)' = x$.

(2) If $x \odot y$ is defined, then $x' \leq y$, so $y' \leq x$, and $(x' + y')' = (y' + x')'$, i.e., $y \odot x$ is defined. Thus, the exchange law is established.

(3) Suppose $y \odot z$, $x \odot (y \odot z)$ are defined, we have $y' \leq z$ and $x' \leq (y' + z')'$. Applying Theorem 2.2 (3), $y' + (y' + z')' = z$. By Theorem 2.2 (1), $y' \leq (y' + z')''$. On the other hand, we get

$(y' + z')'' \leq x''$. Thus, $y' \leq (y' + z')'' \leq x''$, that is, $x' \leq y$. Moreover, above we get

$$(x' + y')'' = y' + x' \leq y' + (y' + z')' = z.$$

Hence, we have

$$(x \odot y) \odot z = (x' + y')'' + z' = x' + (y' + z')'' = x \odot (y \odot z).$$

Thus, the associative law is established.

(4) For any $x, y, h, k \in E$, if $x \leq y$, $h \leq k$, and $x \odot h$, $y \odot k$ are defined, then $y' \leq x'$, $k' \leq h'$, $x' \leq h$, $y' \leq k$. Applying Theorem 2.2 (2),

$$y' + k' \leq x' + k' \leq x' + h', (x' + h')' \leq (y' + k')', x \odot h \leq y \odot k.$$

Therefore, $\odot$ is a partial t-norm.

**Proposition 2.14** Assume $\boldsymbol{C} = (C, \vee, \wedge, \odot, \rightarrow, 0, 1)$ is a commutative quasiresiduated lattice. Then $\odot$ is a partial t-norm on *C*.

**Proof.** If $x \leq y$, $x \odot z$ and $y \odot z$ are defined, then applying Theorem 2.4 (2), $y \leq z \rightarrow (y \odot z)$. Thus $x \leq z \rightarrow (y \odot z)$. Moreover, have $y \odot z \leq z$ by Theorem 2.4 (1) and (3), $x \odot z \leq y \odot z$. Then when $x \leq y$, $h \leq k$, and $x \odot h$, $y \odot k$ are defined, then $x \odot h \leq y \odot h \leq y \odot k$.

Therefore, $\odot$ is a partial t-norm on *C*.

## 3. Partial residuated implications (PRIs) derived from partial t-norms

There are many researchers on the residuated implication induced by t-norms. Therefore, it is necessary to study partial t-norms. In [5], Borzooei gives the concept of a partial t-norm, but has not explained its induced residuated implication. In this section, we will study the residuated implication derived by partial t-norms and call it partial residuated implications.

**Definition 3.1** Assume *L* is a bounded lattice, $\odot$ is a partial t-norm, define:

$$a \rightarrow_{\odot} b := \begin{cases} sup\ \{x \mid a \odot x\ is\ defined\ and\ a \odot x \leq b\}, & if\ the\ supermum\ of\ the\ S\ exists \\ undefined, & otherwise \end{cases}$$

where $S = \{x \mid a \odot x\ is\ defined\ and\ a \odot x \leq b\}$. We call the partial operation $\rightarrow_{\odot}$ on *L* is the partial residuated implication derived by partial t-norm $\odot$.

**Example 3.2** Assume *E* is an LEA. Then Sasaki arrow $\rightarrow_S$ is a PRI on *E*.

**Example 3.3** Assume *E* is an LEA. Define a function $I_S$ as follows:

$$I_S(x, y) := \begin{cases} 1, & if\ x \leq y \\ a', & if\ the\ interval\ E[0, x]\ is\ totally\ ordered, \\ & \quad has\ an\ atom\ a\ and\ x - (x \wedge y) = a \\ 0, & otherwise \end{cases}$$

Then $I_S$ is a partial residuated implication on *E*.

**Example 3.4** Let *L* = {0,1,2,3,4}, the induced order ≤ is showed in Fig. 3. A partial t-norm $\odot$ on *L* and a partial residuated implication $\rightarrow_{\odot}$ derived from $\odot$ are defined by Table 3 and Table 4. Then when $1 \rightarrow_{\odot} 2$ is defined, $sup \neq max$ (see Definition 3.1, since $1 \rightarrow_{\odot} 2 = 4$, but $max\{1,2\}$ does not exist.).

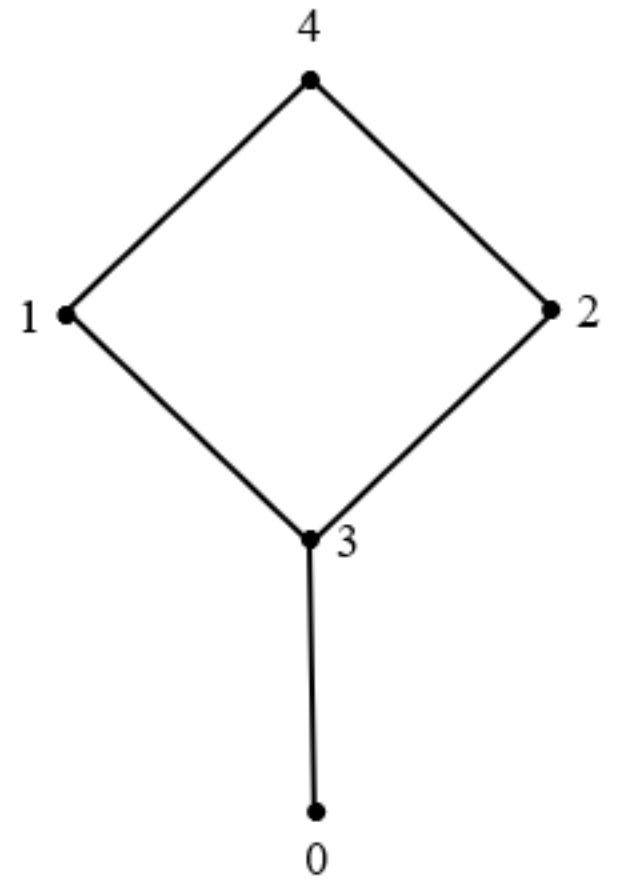

Fig. 3. Order relation

Table 3. Partial operation $\odot$

| $\odot$ | 0 | 1 | 2 | 3 | 4 |
|---|---|---|---|---|---|
| 0 | | | | | 0 |
| 1 | | 3 | 3 | 2 | 1 |
| 2 | | 3 | 3 | 2 | 2 |
| 3 | | 2 | 2 | 3 | 3 |
| 4 | 0 | 1 | 2 | 3 | 4 |

Table 4. Partial operation $\rightarrow_{\odot}$

| $\rightarrow_{\odot}$ | 0 | 1 | 2 | 3 | 4 |
|---|---|---|---|---|---|
| 0 | 4 | 4 | 4 | 4 | 4 |
| 1 | | 4 | 4 | 4 | 4 |
| 2 | | 4 | 4 | 4 | 4 |
| 3 | | 4 | 4 | 4 | 4 |
| 4 | 0 | 1 | 2 | 3 | 4 |

**Theorem 3.5** Assume $L$ is a bounded lattice, $\odot$ is a partial t-norm, $\rightarrow_{\odot}$ is a PRI derived from $\odot$. The following hold:

(i) $\odot$ is infinitely $\vee$-distributive, i.e., if $\bigvee_{i\in I} x_i$ and $\bigvee_{i\in I}(x\odot x_i)$ are existed, then $x\odot(\bigvee_{i\in I} x_i) = \bigvee_{i\in I}(x\odot x_i)$;

(ii) If $x\odot z$ is defined, $x\odot z \leq y$ iff $x \rightarrow_{\odot} y$ is defined and $z \leq x \rightarrow_{\odot} y$;

(iii) If $x \rightarrow_{\odot} y$ and $x\odot(x \rightarrow_{\odot} y)$ are defined, then $x\odot(x \rightarrow_{\odot} y) \leq y$;

(iv) If $\{\, a \mid x\odot a$ is defined and $x\odot a \leq y\}$ is not empty, then the set has the maximum element.

**Proof.** (i) ⇒ (ii): If $x\odot z$ is defined and $x\odot z \le y$, then $z \in \{ a \mid x\odot a$ is defined and $x\odot a \le y\}$, hence $x \to_\odot y = \vee \{ a \mid x\odot a$ is defined and $x\odot a \le y\}$, so $z \le x \to_\odot y$. Conversely, if $z \le x \to_\odot y$, from Definition 2.5 (4), we obtain $x\odot z \le x\odot(x \to_\odot y) = x\odot(\vee \{ a \mid x\odot a$ is defined and $x\odot a \le y\}) = \vee \{x\odot a \mid x\odot a$ is defined and $x\odot a \le y\}) = y$.

(ii) ⇒ (iii): We know $x \to_\odot y \le x \to_\odot y$, then $x\odot(x \to_\odot y) \le y$.

(iii) ⇒ (iv): If $x\odot(x \to_\odot y) \le y$ and $\{ a \mid x\odot a$ is defined and $x\odot a \le y\}$ is a nonempty set , then $x \to_\odot y \in \{ a \mid x\odot a$ is defined and $x\odot a \le y\}$, hence, the set has the maximum element.

(iv) ⇒ (i): If $\{ a \mid x\odot a$ is defined and $x\odot a \le y\}$ is a nonempty set, let $x_i \in \{ a \mid x\odot a$ is defined and $x\odot a \le y\}$, from Definition 2.5 (4), we know $\vee_{i\in I}(x\odot x_i) \le x\odot(\vee_{i\in I} x_i)$. Next, we only need to prove $x\odot(\vee_{i\in I} x_i) \le \vee_{i\in I}(x\odot x_i)$. Let $u = \vee_{i\in I}(x\odot x_i)$, then $x\odot x_i \le u$, we have $x_i \in \{ a \mid x\odot a$ is defined and $x\odot a \le u\}$ for every $x_i \in L$, hence $x_i \le x \to_\odot u$, and $\vee_{i\in I} x_i \le x \to_\odot u$, so $x\odot(\vee_{i\in I} x_i) \le x\odot(x \to_\odot u) \le u$, hence, $x\odot(\vee_{i\in I} x_i) \le \vee_{i\in I}(x\odot x_i)$. In conclusion, $x\odot(\vee_{i\in I} x_i) = \vee_{i\in I}(x\odot x_i)$.

**Corollary 3.6** Assume $L$ is a bounded lattice and $\odot$ is a partial t-norm. Then $\to_\odot$ is a fuzzy implication, if $\odot$ is infinitely $\vee$-distributive, and $\{ a \mid x\odot a$ is defined and $x\odot a \le y\}$ is a nonempty set.

**Proof.** It follows from Theorem 3.5.

We know that in [15] and quasiresiduated lattice, $\to_s$, $\to_d$ and $\to$ are full operations, that is because $\odot$ satisfies Theorem 3.5(ii).

**Theorem 3.7** Assume $L$ is a bounded lattice and $\to_\odot$ is a partial residuated implication derived from a partial t-norm $\odot$ in Definition 3.1. For all $a, b \in L$, when $a \to_\odot b$ is defined, if $a \le b$, then $a \to_\odot b = 1$, the converse is not real.

**Proof.** We know $x \le 1$ and $a\odot 1 = a$, then $a\odot 1 \le b$, hence $a \to_\odot b = 1$. Conversely, by Example 3.4, we know that $1 \to_\odot 2 = 4$, but 1 and 2 cannot compare. So, the converse is not true.

## 4. Partial fuzzy implications (PFIs) and Partial residuated lattices (PRLs)

In this part, we propose the definition of partial fuzzy implications, and explain that the partial residuated implication is a partial fuzzy implication. Then, by defining partial adjoint pairs, we define partial residuated lattices, which are partial algebraic structures corresponding to partial t-norms and partial residuated implications. Finally, the related properties of partial residuated lattices are studied.

**Definition 4.1**[15] A mapping $I: L \times L \to L$ is called a fuzzy implication, if,

(i) $x_1 \le x_2$ implies $I(x_2, y) \le I(x_1, y)$;

(ii) $y_1 \le y_2$ implies $I(x, y_1) \le I(x, y_2)$;

(iii) $I(0,0) = I(1,1) = 1, I(1,0) = 0$.

**Definition 4.2**[19] A mapping $N: L \to L$ is called a negation, if,

(i) $N(0) = 1$ and $N(1) = 0$.

(ii) $x \le y$ implies $N(y) \le N(x)$.

**Definition 4.3** A function $PI: L \times L \to L$ is called a PFI, if,

(PI1) If $x_1 \le x_2$, $PI(x_1, y)$ and $PI(x_2, y)$ are defined, then $PI(x_2, y) \le PI(x_1, y)$;

(PI2) If $y_1 \le y_2$, $PI(x, y_1)$ and $PI(x, y_2)$ are defined, then $PI(x, y_1) \le PI(x, y_2)$;

(PI3) $PI(0,0) = PI(1,1) = 1, PI(1,0) = 0$.

**Example 4.4** Every partial fuzzy implication $PI_N$ can be generated as follows:

$$PI_N(u,v) := \begin{cases} PI\big(N(v), N(u)\big), & if\ PI\big(N(v), N(u)\big)\ is\ defined \\ undefined, & otherwise \end{cases}$$

is called the $N$-reciprocation of a partial fuzzy implication $PI$, where $N$ is a negation and $PI$ is a PFI.

**Example 4.5** Every partial fuzzy implication $PI_N^m$ can be generated as follows:

$$PI_N^m(x,y) := \begin{cases} min\{PI(x,y) \vee N(x), PI_N(x,y) \vee y\}, & if\ PI(x,y)\ and\ PI_N(x,y)\ are\ defined \\ undefined, & otherwise \end{cases}$$

where $PI$ and $PI_N$ are partial fuzzy implications and $N$ is a negation.

**Example 4.6** If $N(u) \in [0, a]$, for all $a < u$, then the following partial fuzzy implication $PI_{I,N}^1$ can be generated:

$$PI_{I,N}^{1v}(x,y) = \begin{cases} 1, & if\ x \le e \\ A, & if\ A\ is\ defined\ and\ e < x \\ undefined, & otherwise \end{cases}$$

Where $a \in (0,1)$, $A = PI\left(N\left(\frac{N(x)}{e}\right), y\right)$, $PI$ is a PFI and $N$ is a negation implication.

**Example 4.7** The following partial fuzzy implication $PI_{PI_1 - PI_2}$ can be generated:

$$PI_{PI_1 - PI_2}(u,v) = \begin{cases} 1, & if\ u = 0 \\ a \cdot PI_1\left(u, \frac{v}{a}\right), & if\ PI_1\ is\ defined\ and\ u > 0, v \le a \\ a + (1-a) \cdot PI_2\left(u, \frac{v-a}{1-a}\right), & if\ PI_2\ is\ defined\ and\ u > 0, u > a \\ undefined, & otherwise \end{cases}$$

where $PI_1$ and $PI_2$ are two partial fuzzy implications and $a \in (0,1)$.

**Theorem 4.8** Assume $\odot$ is a partial t-norm on bounded lattice $L$ and $\to_\odot$ is a PRI derived from $\odot$. Then $\to_\odot$ must be the partial fuzzy implication.

**Proof.** (PI1) If $a \to_\odot c$ and $b \to_\odot c$ are defined, then $a \to_\odot c = sup\{x_1 \mid a \odot x_1$ is defined and $a \odot x_1 \le c\}$, $b \to_\odot c = sup\{x_2 \mid b \odot x_2$ is defined and $b \odot x_2 \le c\}$, i.e., $\exists x_2$, s. t., $b \odot x_2$ is defined an d $b \odot x_2 \le c$, hence $b \le x_2 \to_\odot c$. And when $a \le b$, then $a \le x_2 \to_\odot c$, so $a \odot x_2$ is defined and $a \odot x_2 \le c$, then $x_2 \in \{x_1 \mid a \odot x_1$ is defined and $a \odot x_1 \le c\}$, and $\{x_2 \mid b \odot x_2$ is defined and $b \odot x_2 \le c\} \subseteq \{x_1 \mid a \odot x_1$ is defined and $a \odot x_1 \le c\}$, hence $sup\{x_2 \mid b \odot x_2$ is defined and $b \odot x_2 \le c\} \subseteq sup\{x_1 \mid a \odot x_1$ is defined and $a \odot x_1 \le c\}$, i.e., $b \to_\odot c \le a \to_\odot c$.

(PI2) Similar to (PI1), we can get $a \to_\odot b \le a \to_\odot c$.

(PI3) $0 \rightarrow_{\odot} 0 = sup\{a \mid 0 \odot a$ is defined and $0 \odot a \leq 0\} = sup\{a \mid 0 \odot a$ is defined and $0 \odot a = 0\} = 1$, i.e., $PI(0,0) = 1$;

$1 \rightarrow_{\odot} 1 = sup\{a \mid 1 \odot a$ is defined and $1 \odot a \leq 1\} = 1$, i.e., $PI(1,1) = 1$;

$1 \rightarrow_{\odot} 0 = sup\{a \mid 1 \odot a$ is defined and $1 \odot a \leq 0\} = sup\{a \mid 1 \odot a$ is defined and $1 \odot a = 0\} = 0$, i.e., $PI(1,0) = 0$.

**Definition 4.9** The operations $\otimes$ and $\rightarrow$ are two partial operations. A pair $(\otimes, \rightarrow)$ on a poset $(P; \leq)$, is called a partial adjoint pair, if the following (PAP1), (PAP2), (PAP3) are hold,

(PAP1) The operation $\otimes$ is isotone, if $x \leq y$, $x \otimes z$ and $y \otimes z$ are defined, then $x \otimes z \leq y \otimes z$; if $x \leq y$, $z \otimes x$ and $z \otimes y$ are defined, then $z \otimes x \leq z \otimes y$.

(PAP2) The operation $\rightarrow$ is antitone in the first argument, if $x \leq y$, $x \rightarrow z$ and $y \rightarrow z$ are defined, then $y \rightarrow z \leq x \rightarrow z$; $\rightarrow$ is isotone in the second argument, if $x \leq y$, $z \rightarrow x$ and $z \rightarrow y$ are defined, then $z \rightarrow x \leq z \rightarrow y$.

(PAP3) If $x \otimes y$ and $x \rightarrow z$ are defined, then $x \otimes y \leq z$ iff $y \leq x \rightarrow z$.

**Definition 4.10** A partial residuated lattice $(L; \leq, \otimes, \rightarrow, 0, 1)$, where $(L; \leq, 0, 1)$ is a bounded lattice, $\otimes$ and $\rightarrow$ are two partial operations, if it satisfies,

(PRL1) If $x \otimes y$ is defined, then $y \otimes x$ is defined, $x \otimes y = y \otimes x$;

(PRL2) If $y \otimes z$, $x \otimes (y \otimes z)$ are defined, then $x \otimes y$, $(x \otimes y) \otimes z$ are defined, and $x \otimes (y \otimes z) = (x \otimes y) \otimes z$;

(PRL3) $x \otimes 1$ is defined and $x \otimes 1 = x$;

(PRL4) $(\otimes, \rightarrow)$ is a partial adjoint pair on $L$.

**Example 4.11** If $L = \{0,1,2,3\}$, the induced order $\leq$ is depicted in Fig. 4. The operations $\otimes$ and $\rightarrow$ are defined by Table 5 and Table 6. Then $L$ is a PRL.

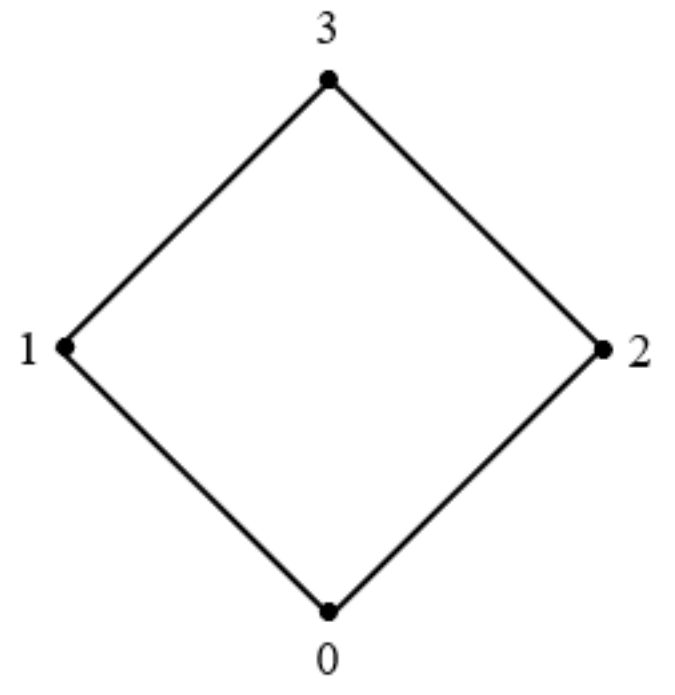

Fig. 4. Order relation

Table 5. Partial operation $\otimes$

| $\otimes$ | 0 | 1 | 2 | 3 |
|---|---|---|---|---|
| 0 | | | 0 | 0 |
| 1 | | 1 | | 1 |
| 2 | 0 | | 2 | 2 |
| 3 | 0 | 1 | 2 | 3 |

Table 6. Partial operation →

| → | 0 | 1 | 2 | 3 |
|---|---|---|---|---|
| 0 | 3 | | | |
| 1 | 2 | 3 | | |
| 2 | 0 | | | |
| 3 | 0 | 1 | | |

**Example 4.12** If $L$ = {0,1,2,3,4}, the induced order $\leq$ is showed in Fig. 1. The operations $\otimes$ and → are defined by Table 7 and Table 8. Then $L$ is a PRL.

Table 7. Partial operation $\otimes$

| $\otimes$ | 0 | 1 | 2 | 3 | 4 |
|---|---|---|---|---|---|
| 0 | | | | | 0 |
| 1 | | | | 0 | 1 |
| 2 | | | 2 | | 2 |
| 3 | | 0 | | 3 | 3 |
| 4 | 0 | 1 | 2 | 3 | 4 |

Table 8. Partial operation →

| → | 0 | 1 | 2 | 3 | 4 |
|---|---|---|---|---|---|
| 0 | 4 | 4 | 4 | 4 | 4 |
| 1 | 3 | 4 | 4 | | 4 |
| 2 | 3 | | 4 | | 4 |
| 3 | 2 | 2 | | 4 | 4 |
| 4 | 0 | 1 | 2 | 3 | 4 |

**Theorem 4.13** Assume $(L; \leq, \otimes, \rightarrow, 0, 1)$ is a PRL. Then,

(1) $x \rightarrow x$ is defined, implies $x \rightarrow x = 1$.

(2) $x \rightarrow 1$ is defined, implies $x \rightarrow 1 = 1$.

(3) $1 \rightarrow x$ is defined, implies $1 \rightarrow x = x$.

(4) $x \rightarrow y$ is defined, implies $x \rightarrow y = 1$ iff $x \leq y$.

**Proof.** (1) We know $x \otimes 1 \leq x$, by (PAP3), we get $1 \leq x \rightarrow x$, then $x \rightarrow x = 1$.

(2) We know $x \otimes 1 \leq 1$, if $x \rightarrow 1$ is defined, then $1 \leq x \rightarrow 1$, further $x \rightarrow 1 = 1$.

(3) Since $1 \otimes x \leq x$, so $x \leq 1 \to x$. And since $1 \to x \leq 1 \to x$, then $(1 \to x) \otimes 1 \leq x$, so $1 \to x = x$.

(4) (⇒) For all $x, y \in L$, $1 \leq x \to y$, so $x \otimes 1 \leq y$, hence $x \leq y$.

(⇐) For all $x, y \in L$, $x \otimes 1 \leq y$, so $1 \leq x \to y$, hence $x \to y = 1$.

**Theorem 4.14** Assume $L$ is a bounded lattice and $\to_{\odot}$ is a PRI derived from a partial t-norm $\odot$. Then $(L; \leq, \odot, \to_{\odot}, 0,1)$ is a PRL.

**Proof.** By Definition 2.5 and Definition 3.1, we can easily get that (PAP1), (PAP3), (PRL1), (PRL2) and (PRL3) hold, next, prove (PAP2).

For any $x, y, z \in L$, suppose $x \leq y$, if $a \in L$, $z \odot a$ is defined and $z \odot a \leq x$, then $z \odot a \leq x \leq y$. That is $\{a \in L \mid z \odot a \text{ is defined and } z \odot a \leq x\} \subseteq \{b \in L \mid z \odot b \text{ is defined and } z \odot b \leq y\}$, hence $\sup\{a \in L \mid z \odot a \text{ is defined and } z \odot a \leq x\} \leq \sup\{b \in L \mid z \odot b \text{ is defined and } z \odot b \leq y\}$ . So $z \to_{\odot} x \leq z \to_{\odot} y$. Samely, we can get $y \to_{\odot} z \leq x \to_{\odot} z$.

**Theorem 4.15** If $E = (E, \leq, +, ', 0, 1)$ is an LEA, and,

$$x \odot y := (x' + y')' \text{ iff } x' \leq y.$$

$$x \to y := x' + y \text{ iff } y \leq x.$$

Then $E$ is a PRL.

**Proof.** It is follows from Proposition 2.13, that $\odot$ is a partial t-norm, then (PRL1), (PRL2) and (PRL3) hold, we only need to prove (PRL4). It is obvious that (PAP1) holds, next, we will prove (PAP2) and (PAP3).

(PAP2) On the one hand, if $x \leq y$, then $y' \leq x'$. And $x \to z = x' + z$, $y \to z = y' + z$, hence, $y' + z \leq x' + z$, $y \to z \leq x \to z$. On the other hand, we can get similar results: $x \to y \leq x \to z$.

(PAP3) First of all, we know if $x \odot y \leq z$, then $(x' + y')' \leq z$, hence $z' \leq x' + y'$. In other words, there exists $u \in E$, $u + z' = x' + y'$, so $(u + z')' \leq z$. From the properties of lattice effect algebra, $y' = (x' + (u + z')')' \Leftrightarrow y = x' + (u + z')'$, so $x' + (u + z')' \leq x' + z$. Thus, $y \leq x' + z \Leftrightarrow y \leq x \to z$. This means the residuated condition is true.

**Definition 4.16** The operations $\otimes$ and $\to$ are two partial operations. A pair $(\otimes, \to)$ on a poset $(P; \leq)$, is called a special partial adjoint pair (sPAP), if the following (sPAP1), (sPAP2), (sPAP3) hold,

(sPAP1) The operation $\otimes$ is isotone, if $x \leq y$ and $x \otimes z$ is defined, then $y \otimes z$ is defined, $x \otimes z \leq y \otimes z$; if $x \leq y$ and $z \otimes x$ is defined, then $z \otimes y$ is defined, $z \otimes x \leq z \otimes y$.

(sPAP2) The operation $\to$ is antitone in the first argument, if $x \leq y$ and $x \to z$ is defined, then $y \to z$ is defined, $y \to z \leq x \to z$; and $\to$ is isotone in the second argument, if $x \leq y$ and $z \to y$ is defined, then $z \to x$ is defined, $z \to x \leq z \to y$.

(sPAP3) If $x \otimes y$ is defined and $x \otimes y \leq z$ iff $x \to z$ is defined and $y \leq x \to z$.

**Definition 4.17** A special partial residuated lattice $(L; \leq, \otimes, \to, 0,1)$, where $(L; \leq, 0,1)$ is a bounded lattice, $\otimes$ and $\to$ are two partial operations, if,

(sPRL1) If $x \otimes y$ is defined, then $y \otimes x$ is defined and $x \otimes y = y \otimes x$;

(sPRL2) If $y \otimes z$ and $x \otimes (y \otimes z)$ are defined, then $x \otimes y$ and $(x \otimes y) \otimes z$ are defined and $x \otimes (y \otimes z) = (x \otimes y) \otimes z$;

(sPRL3) $1 \otimes x$ is defined and $1 \otimes x = x$;

(sPRL4) $(\otimes, \rightarrow)$ is an sPAP.

We abbreviate special partial residuated lattice as sPRL.

**Theorem 4.18** If $L = (L; \leq, \otimes, \rightarrow, 0,1)$ is an sPRL. Then $L$ is a residuated lattice.

**Proof.** (1) For all $x \in L$, $x \otimes 1$ is defined, $x \otimes 1 \leq x$, so $1 \leq x \rightarrow x$, further, $x \rightarrow x = 1$.

(2) For all $x \in L$, we have $x \leq 1 = 0 \rightarrow 0$, so $x \otimes 0$ is defined and $x \otimes 0 \leq 0$, so $x \otimes 0 = 0$.

(3) By (2), we know $x \otimes 0 = 0$, so $x \otimes 0 \leq y$, then $x \rightarrow y$ is defined and $0 \leq x \rightarrow y$.

(4) By (1), we know $x \leq 1 = y \rightarrow y$, so $x \otimes y$ is defined and $x \otimes y \leq y$.

To sum up, $L$ is a residuated lattice.

**Definition 4.19** Assume $L = (L; \leq, \otimes, \rightarrow, 0,1)$ is a PRL. We call $L$ is well (wPRL). For all $x, y \in L$, if it satisfies,

(W1) $x \rightarrow x$ is defined, $x \rightarrow 1$ is defined;

(W2) $x \rightarrow y$ is defined, implies $x \otimes (x \rightarrow y)$ is defined.

**Example 4.20** If $L = \{0,1,2,3\}$, the induced order $\leq$ is showed in Fig. 4. The operations $\otimes$, $\rightarrow$ are defined by Table 9 and Table 10. Then $L$ is a wPRL.

Table 9. Partial operation $\otimes$

| $\otimes$ | 0 | 1 | 2 | 3 |
|---|---|---|---|---|
| 0 | | | | 0 |
| 1 | | | 0 | 1 |
| 2 | | 0 | | 2 |
| 3 | 0 | 1 | 2 | 3 |

Table 10. Partial operation $\rightarrow$

| $\rightarrow$ | 0 | 1 | 2 | 3 |
|---|---|---|---|---|
| 0 | 3 | | | 3 |
| 1 | | 3 | | 3 |
| 2 | | 1 | 3 | 3 |
| 3 | 0 | 1 | 2 | 3 |

**Example 4.21** If $L = \{0,1,2,3,4\}$, the induced order $\leq$ is showed in Fig. 1. The operations $\otimes$, $\rightarrow$ are defined by Table 11 and Table 12. Then $L$ is a wPRL.

Table 11. Partial operation ⊗

| ⊗ | 0 | 1 | 2 | 3 | 4 |
|---|---|---|---|---|---|
| 0 | | | | | 0 |
| 1 | | | | 0 | 1 |
| 2 | | | 2 | 0 | 2 |
| 3 | | 0 | 0 | 3 | 3 |
| 4 | 0 | 1 | 2 | 3 | 4 |

Table 12. Partial operation →

| → | 0 | 1 | 2 | 3 | 4 |
|---|---|---|---|---|---|
| 0 | 4 | 4 | 4 | 4 | 4 |
| 1 | 3 | 4 | 4 | | 4 |
| 2 | 3 | | 4 | | 4 |
| 3 | | | | 4 | 4 |
| 4 | 0 | 1 | 2 | 3 | 4 |

**Example 4.22** If $L$ = {0,1,2,3,4,5}, the induced order ≤ is showed in Fig. 5. The operations ⊗, → are defined by Table 13 and Table 14. Then $L$ is a wPRL.

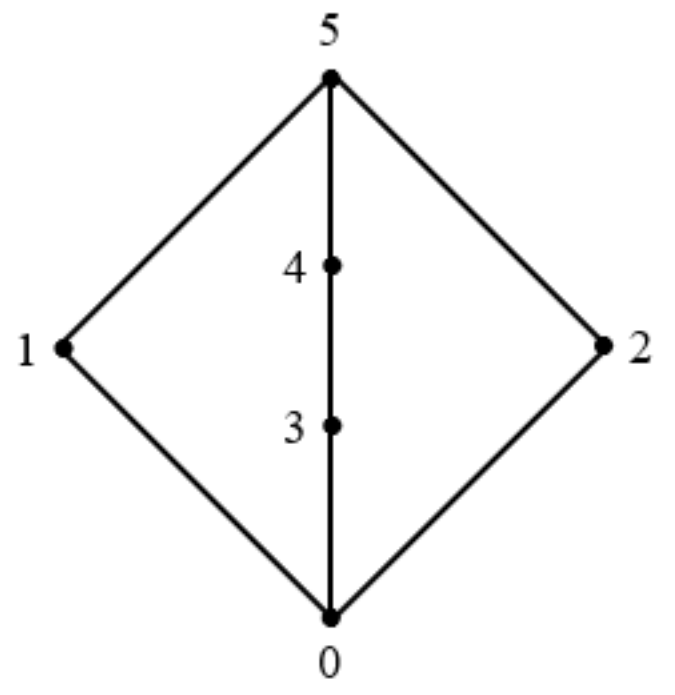

Fig. 5. Order relation

Table 13. Partial operation ⊗

| ⊗ | 0 | 1 | 2 | 3 | 4 | 5 |
|---|---|---|---|---|---|---|
| 0 | 0 | | | 0 | 0 | 0 |
| 1 | | | | | | 1 |
| 2 | | | | | | 2 |
| 3 | 0 | | | 0 | 0 | 3 |
| 4 | 0 | | | 0 | 4 | 4 |
| 5 | 0 | 1 | 2 | 3 | 4 | 5 |

Table 14. Partial operation →

| → | 0 | 1 | 2 | 3 | 4 | 5 |
|---|---|---|---|---|---|---|
| 0 | 5 | 5 | 5 | 5 | 5 | 5 |
| 1 | | 5 | | | | 5 |
| 2 | | | 5 | | | 5 |
| 3 | 4 | | | 5 | 5 | 5 |
| 4 | 3 | | | 3 | 5 | 5 |
| 5 | 0 | 1 | 2 | 3 | 4 | 5 |

**Example 4.23** If $L = \{0,1,2,3,4,5\}$, Fig. 6 shows the partial order $\leq$. The operations $\otimes$, $\rightarrow$ are defined by Table 15 and Table 16. Then $L$ is a wPRL.

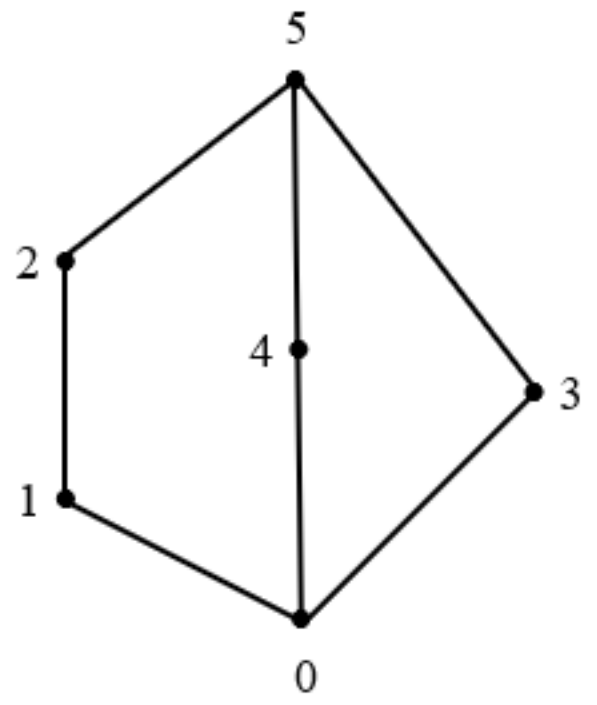

Fig. 6. Order relation

Table 15. Partial operation ⊗

| ⊗ | 0 | 1 | 2 | 3 | 4 | 5 |
|---|---|---|---|---|---|---|
| 0 | | | | | | 0 |
| 1 | | | 0 | | | 1 |
| 2 | | 0 | 2 | | | 2 |
| 3 | 0 | | | 0 | | 3 |
| 4 | 0 | | | | 0 | 4 |
| 5 | 0 | 1 | 2 | 3 | 4 | 5 |

Table 16. Partial operation →

| → | 0 | 1 | 2 | 3 | 4 | 5 |
|---|---|---|---|---|---|---|
| 0 | 5 | 5 | 5 | 5 | 5 | 5 |
| 1 | 2 | 5 | | | | 5 |
| 2 | 1 | 1 | 5 | | | 5 |
| 3 | 3 | | | 5 | | 5 |
| 4 | 4 | | | | 5 | 5 |
| 5 | 0 | 1 | 2 | 3 | 4 | 5 |

**Theorem 4.24** Consider $(L; \leq, \otimes, \rightarrow, 0, 1)$ is a wPRL, then,

(1) If $x \otimes y$ is defined, then $x \otimes y \leq x \wedge y$;

(2) If $x \rightarrow y$ is defined, then $x \otimes (x \rightarrow y) \leq y$;

(3) If $x \rightarrow y$ is defined, then $x \leq (x \rightarrow y) \rightarrow y$.

**Proof.** (1) Since $x \leq 1 = y \rightarrow y$, then $x \otimes y \leq y$; obviously, $x \otimes y \leq x$. Hence, $x \otimes y \leq x \wedge y$.

(2) Based on assumptions and by (W2), $x \otimes (x \rightarrow y)$ is defined, since $x \rightarrow y \leq x \rightarrow y$, by (PAP3), $x \otimes (x \rightarrow y) \leq y$.

(3) Based on assumptions and by (2), $x \otimes (x \rightarrow y) \leq y$, applying (PRL1) and (W2), $(x \rightarrow y) \rightarrow y$ is defined, applying (PAP3), we have $x \leq (x \rightarrow y) \rightarrow y$.

**5. Partial t-conorms and partial co-residuated lattices**

In [25], Zhou and Li further investigate the relationship between residuated structures and some quantum structures from the perspective of partial algebra, and introduce the concept of partial residuated lattice. In order to avoid ambiguity, we call it a $z\mathcal{L}$- partial residuated lattice ($z\mathcal{L}$-PRL). In this section, we introduce PcRL and show that $z\mathcal{L}$-PRL is a PcRL and also a co-residuated lattice.

**Definition 5.1**[14] The operation $\oplus$ is a t-conorm, if,

(i) $0 \oplus x = x$;

(ii) $x \oplus y = y \oplus x$;

(iii) $x \oplus (y \oplus z) = (x \oplus y) \oplus z$;

(iv) if $x \leq y$, $h \leq k$; then $x \oplus h \leq y \oplus k$.

**Definition 5.2**[24] The operations $\oplus$, $\ominus$ are two binary operations. A pair $(\oplus, \ominus)$ is called a co-adjoint pair, the conditions (cAP1), (cAP2), (cAP3) hold,

(cAP1) The operation $\oplus$ is isotone, $x \leq y$ implies $x \oplus z \leq y \oplus z$; $x \leq y$ implies $z \oplus x \leq z \oplus y$.

(cAP2) The operation $\ominus$ is isotone in the first argument, $x \leq y$ implies $x \ominus z \leq y \ominus z$; and $\ominus$ is antitone in the second argument, $x \leq y$ implies $z \ominus y \leq z \ominus x$.

(cAP3) $z \leq x \oplus y$ iff $z \ominus y \leq x$.

**Definition 5.3**[24] A structure $(S; \leq, \oplus, \ominus, 0, 1)$, where $\oplus$, $\ominus$ are two binary operations, it is a co-residuated lattice, if,

(cRL1) $(S; \oplus, 0)$ is a commutative semigroup;

(cRL2) $x \oplus 0 = x$;

(cRL3) $(\oplus, \ominus)$ is a co-adjoint pair.

**Definition 5.4** Let $S$ be a bounded lattice. Then $\circledast$ is a partial operation, it is a partial t-conorm, if,

(i) $0 \circledast x = x$;

(ii) If $x \circledast y$ is defined, then $y \circledast x$ is defined, $x \circledast y = y \circledast x$;

(iii) If $y \circledast z$ and $x \circledast (y \circledast z)$ are defined, then $x \circledast y$ and $(x \circledast y) \circledast z$ are also defined and $x \circledast (y \circledast z) = (x \circledast y) \circledast z$;

(iv) If $x \leq y$, $h \leq k$, $x \circledast h$ and $y \circledast k$ are defined, then $x \circledast h \leq y \circledast k$.

From the above definitions, we can easily conclude that in a bounded lattice, every t-conorm is a partial t-conorm. And we can regard a partial t-conorm as a common generalization of partial t-norm and conorm.

**Example 5.5** Let $S = [0,1]$. Define $\circledast$ as a partial operation:

$$a \circledast b = \begin{cases} undefined, & if\ a, b \in [0.5,1] \\ max\{a, b\}, & others \end{cases}$$

$(a, b \in S)$. It is obviously that $\circledast$ is a partial t-conorm.

**Example 5.6** Let $S = [0,1]$. Define $\circledast$ as a partial operation:

$$a \circledast b = \begin{cases} max\{a, b\}, & if\ a, b \in [0,0.5]\ or\ a = 0\ or\ b = 0 \\ undefined, & others \end{cases}$$

$(a, b \in S)$. Then $\circledast$ is a partial t-conorm.

**Example 5.7** Let $S = [0,1]$. Define $\circledast$ as a partial operation:

$$a \circledast b = \begin{cases} a \vee b, & if\ a + b \leq 1\ or\ a = 0\ or\ b = 0 \\ undefined, & others \end{cases}$$

$(a, b \in S)$. Then $\circledast$ is a partial t-conorm.

**Example 5.8** Let $S = [0,1]$. Define $\circledast$ as a partial operation:

$$a \circledast b = \begin{cases} a \vee b, & if\ a + b \leq 0.5\ or\ a = 0\ or\ b = 0 \\ undefined, & others \end{cases}$$

$(a, b \in S)$. Then $\circledast$ is a partial t-conorm.

**Example 5.9** Let $S = [0,1]$. Define $\circledast$ as a partial operation:

$$a \circledast b = \begin{cases} a \vee b, & if\ a + b \leq \alpha\ or\ a = 0\ or\ b = 0 \\ undefined, & others \end{cases}$$

$(x, y \in S)$, where $\alpha \in [0,1]$. Then $\circledast$ is a partial t-conorm.

**Example 5.10** If $S = \{0,1,2,3,4\}$, Fig. 1 shows the partial order $\leq$. Then $\circledast$ is defined by the following table. Thus $\circledast$ is a partial t-conorm.

Table 17. The operation $\circledast$ on $S$

| $\circledast$ | 0 | 1 | 2 | 3 | 4 |
|---|---|---|---|---|---|
| 0 | 0 | 1 | 2 | 3 | 4 |
| 1 | 1 | 2 | 2 | | |
| 2 | 2 | 2 | 2 | | |
| 3 | 3 | | | 3 | |
| 4 | 4 | | | | |

**Example 5.11** Let $S = \{0,1,2,3,4\}$, Fig. 2 shows the partial order $\leq$. Then $\circledast$ is defined by the following table. Thus $\circledast$ is a partial t-conorm.

Table 18. The operation $\circledast$ on $S$

| $\circledast$ | 0 | 1 | 2 | 3 | 4 |
|---|---|---|---|---|---|
| 0 | 0 | 1 | 2 | 3 | 4 |
| 1 | 1 | 1 | | | |
| 2 | 2 | | 2 | | |
| 3 | 3 | | | 3 | |
| 4 | 4 | | | | |

**Definition 5.12** Let $S$ be a bounded lattice, $\circledast$ be a partial t-conorm, for any $a, b \in S$, define $a \rightarrow_{\circledast} b$ as following:

$$a \rightarrow_{\circledast} b := \begin{cases} inf\{x \mid a \circledast x \text{ is defined and } a \circledast x \geq b\}, & \text{if the infimum of the } I \text{ exists} \\ undefined, & otherwise \end{cases}$$

where $I = \{x \mid a \circledast x \text{ is defined and } a \circledast x \geq b\}$. We call the partial operation $\rightarrow_{\circledast}$ on $S$ is the partial residuated co-implication derived by partial t-conorm $\circledast$.

**Definition 5.13** The operations $\circledast$, $\rightsquigarrow$ are two partial operations, a pair $(\circledast, \rightsquigarrow)$ on a poset $(P; \leq)$ is called a partial co-adjoint pair, if the following properties (cPAP1), (cPAP2), (cPAP3) are satisfied,

(cPAP1) The operation $\circledast$ is isotone, if $x \leq y$, $x \circledast z$ and $y \circledast z$ are defined, then $x \circledast z \leq y \circledast z$.

(cPAP2) The operation $\rightsquigarrow$ is isotone in the first argument, if $x \leq y$, $x \rightsquigarrow z$ and $y \rightsquigarrow z$ are defined, then $x \rightsquigarrow z \leq y \rightsquigarrow z$; and $\rightsquigarrow$ is antitone in the second variable, if $x \leq y$, $z \rightsquigarrow y$ and $z \rightsquigarrow x$ are defined, then $z \rightsquigarrow y \leq z \rightsquigarrow x$.

(cPAP3) If $x \circledast y$ and $z \rightsquigarrow y$ are defined, then $z \leq x \circledast y$ iff $z \rightsquigarrow y \leq x$.

**Definition 5.14** A structure $(S; \leq, \circledast, \rightsquigarrow, 0, 1)$ is called a partial co-residuated lattice (PcRL), where $(S; \leq, 0, 1)$ is a bounded lattice, $\circledast$ and $\rightsquigarrow$ are two partial operations, if it satisfies,

(cPRL1) If $x \circledast y$ is defined, then $y \circledast x$ is defined and $x \circledast y = y \circledast x$;

(cPRL2) If $y \circledast z$, $x \circledast (y \circledast z)$ are defined, then $x \circledast y$, $(x \circledast y) \circledast z$ are defined and $x \circledast (y \circledast z) = (x \circledast y) \circledast z$;

(cPRL3) $x \circledast 0$ is defined and $x \circledast 0 = x$;

(cPRL4) $(\circledast, \rightsquigarrow)$ is a partial co-adjoint pair on $S$.

Actually, it is to be noted here that the partial co-residuated lattice given above is a dual concept by reversing the order of a PRL.

**Example 5.15** If $S = \{0,1,2,3\}$, the induced order $\leq$ is showed in Fig. 4. The operations $\circledast$, $\rightsquigarrow$ are defined by Table 19 and Table 20. Then $S$ is a PcRL.

Table 19. The operation $\circledast$ on $S$

| $\circledast$ | 0 | 1 | 2 | 3 |
|---|---|---|---|---|
| 0 | 0 | 1 | 2 | 3 |
| 1 | 1 | | | |
| 2 | 2 | | | |
| 3 | 3 | | | 3 |

Table 20. The operation $\rightsquigarrow$ on $S$

| $\rightsquigarrow$ | 0 | 1 | 2 | 3 |
|---|---|---|---|---|
| 0 | 0 | | | |
| 1 | | 1 | 3 | 1 |
| 2 | | 3 | 2 | 2 |
| 3 | | | | 3 |

**Example 5.16** Let $S = \{0,1,2,3,4\}$, Fig. 1 shows the partial order $\leq$. The operations $\circledast$, $\rightsquigarrow$ are defined by Table 21 and Table 22. Then $S$ is a PcRL.

Table 21. The operation $\circledast$ on $S$

| $\circledast$ | 0 | 1 | 2 | 3 | 4 |
|---|---|---|---|---|---|
| 0 | 0 | 1 | 2 | 3 | 4 |
| 1 | 1 | | | | |
| 2 | 2 | | | | |
| 3 | 3 | | | | |
| 4 | 4 | | | | |

Table 22. The operation $\rightsquigarrow$ on $S$

| $\rightsquigarrow$ | 0 | 1 | 2 | 3 | 4 |
|---|---|---|---|---|---|
| 0 | 0 | 0 | 0 | 0 | 0 |
| 1 | | 1 | | | |
| 2 | | | 2 | | 2 |
| 3 | | 2 | 2 | 3 | 0 |
| 4 | | | | | 4 |

**Theorem 5.17** Let $(S;\leq,\circledast,\rightsquigarrow,0,1)$ be a PcRL, then,

(1) If $x \rightsquigarrow 0$ is defined, then $x \rightsquigarrow 0 = x$.

(2) If $x \rightsquigarrow y$ is defined, then $x \rightsquigarrow y = 0$ iff $x \leq y$.

(3) If $x \circledast y$ and $(x \circledast y) \rightsquigarrow y$ are defined, then $(x \circledast y) \rightsquigarrow y \leq x$.

(4) If $x \rightsquigarrow y$ and $(x \rightsquigarrow y) \circledast y$ are defined, then $x \leq (x \rightsquigarrow y) \circledast y$.

**Proof.** (1) Because $x \leq x \circledast 0$, then $x \rightsquigarrow 0 \leq x$. And for any $a \in S$, $x \rightsquigarrow 0 \leq a$, $x \leq a \circledast 0 = a$. Let $a = x \rightsquigarrow 0$, we get $x \leq x \rightsquigarrow 0$. So, $x \rightsquigarrow 0 = x$.

(2) ($\Rightarrow$) From the conditions, since $x \rightsquigarrow y \leq x \rightsquigarrow y$ and $0 \leq x \rightsquigarrow y$, so $x \circledast 0 \leq y$, hence $x \leq y$.
($\Leftarrow$) We have $x \leq y \circledast 0$, then $x \rightsquigarrow y \leq 0$, so, $x \rightsquigarrow y = 0$.
(3) We know $x \circledast y \leq x \circledast y$, applying (cPAP3), $(x \circledast y) \rightsquigarrow y \leq x$.
(4) We know $x \rightsquigarrow y \leq x \rightsquigarrow y$, applying (cPAP3), $x \leq (x \rightsquigarrow y) \circledast y$.
**Definition 5.18**[25] A structure $(S;\leq,\oplus,\ominus,0,1)$ is called a partial residuated lattice, $\oplus$ and $\ominus$ are two partial operations, if:

(i) $(S;\leq)$ is a bounded lattice;

(ii) $(S;\oplus,0)$ is a partial commutative monoid, its unit element is 0;

(iii) $(\oplus,\ominus)$ is a partial adjoint pair.

**Note:** The partial residuated lattice in Definition 5.18 is a $z\mathcal{L}$-PRL.

Next, we will first prove that the $z\mathcal{L}$-PRL is a partial co-residuated lattice, and then prove that it is a co-residuated lattice. In other words, we show that the $\oplus$ and $\ominus$ operations in $z\mathcal{L}$-PRL are full operations.

**Theorem 5.19** Let $(S;\leq,\oplus,\ominus,0,1)$ be a $z\mathcal{L}$-PRL, if,

$$a \leq b \triangleq b \preccurlyeq a,$$

$$\imath \triangleq 0,$$

$$\theta \triangleq 1.$$

then $(S;\preccurlyeq,\oplus,\ominus,\theta,\imath)$ is a partial co-residuated lattice.

**Proof.** Obviously, we know that if we want it to be a partial co-residuated lattice, we only need to explain that (cPAP3) and (cPRL3) are true. Under the condition of meaning, we have:

(1) If $x \oplus y$ and $z \ominus y$ are defined, then $z \preccurlyeq x \oplus y$ iff $z \ominus y \preccurlyeq x$.

(2) $x \oplus \imath$ is defined and $x \oplus \imath = x$;

It is easy to get $(S;\preccurlyeq,\oplus,\ominus,\theta,\imath)$ is a PcRL.

**Corollary 5.20** If $(S; \preccurlyeq, \oplus, \ominus, \theta, \imath)$ is a PcRL. Then it is a co-residuated lattice.

**Proof.** It can be proved by Theorem 4.18 and Theorem 5.19.

## 6. Filters in well partial residuated lattices (wPRLs)

In this section, we define filters and give strong filters on well partial residuated lattices. Finally, the quotient structure $(L/{\sim_F}; \leq, \otimes, \rightarrow, [0]_F, [1]_F)$ is constructed, and it is proved that $(L/{\sim_F}; \leq, \otimes, \rightarrow, [0]_F, [1]_F)$ is an PRL.

**Definition 6.1** Assume $(L; \leq, \otimes, \rightarrow, 0,1)$ is a wPRL. $F \neq \emptyset$, which is called a filter, if,

(F1) $1 \in F$;

(F2) If $x \in F$, $y \in L$, $x \leq y$, then $y \in F$;

(F3) If $x \in F$, $y \in F$, $x \otimes y$ is defined, then $x \otimes y \in F$.

If $F \neq L$, then filter $F$ is called proper.

**Example 6.2** In Example 4.20, $L = \{0,1,2,3\}$ and $(L; \leq, \otimes, \rightarrow, 0,1)$ is a wPRL, the proper filters are: $\{3\}$, $\{1,3\}$, $\{2,3\}$.

**Example 6.3** In Example 4.21, $L = \{0,1,2,3,4\}$ and $(L; \leq, \otimes, \rightarrow, 0,1)$ is a wPRL, the proper filters are: $\{4\}$, $\{2,4\}$, $\{3,4\}$, $\{1,2,4\}$.

**Example 6.4** In Example 4.22, $L = \{0,1,2,3,4,5\}$ and $(L; \leq, \otimes, \rightarrow, 0,1)$ is a wPRL, the proper filters are: $\{5\}$, $\{1,5\}$, $\{2,5\}$, $\{4,5\}$, $\{1,2,5\}$, $\{1,4,5\}$, $\{2,4,5\}$.

**Example 6.5** Let $L = [0,1]$. Define $\otimes$ and $\rightarrow$ are two partial operations:

$$a \otimes b = \begin{cases} undefined, & if\ a, b \in [0,0.5] \\ min\{a, b\}, & others \end{cases}$$

$$a \rightarrow b = \begin{cases} 1, & a \leq b \\ b, & a > b \end{cases}$$

$(a, b \in L)$. Then $(L; \leq, \otimes, \rightarrow, 0,1)$ is a PRL, the proper filters are: $F1 = \{1\}$, $F2 = [x, 1]$, where $x \in [0,1]$.

**Example 6.6** Let $L = [0,1]$. Define $\otimes$ and $\rightarrow$ are two partial operations:

$$a \otimes b = \begin{cases} a \wedge b, & if\ a + b \leq 1\ or\ a = 1\ or\ b = 1 \\ undefined, & others \end{cases}$$

$$a \rightarrow b = \begin{cases} 1, & a \leq b \\ (1 - a) \vee b, & a > b \end{cases}$$

$(a, b \in L)$. Then $(L; \leq, \otimes, \rightarrow, 0,1)$ is a PRL, the proper filters are: $\{1\}$ (for any $x \in [0,1]$, $[x, 1]$ is not a filter in $L$, if $x = 0.3$, then $0.3 \rightarrow 0.1 = (1 - 0.3) \vee 0.1 = 0.7 \in F$, $0.1 \in F$, but $0.1 \notin [0.3,1]$. It is similarly that $[1 - x, 1]$ is not a filter in $L$, for any $x \in [0,1]$, if $x = 0.3$, then $0.3 \rightarrow 0.4 = 1 \in F$, $0.4 \in F$, but $0.4 \notin [0.7,1]$).

**Example 6.7** Let $L = [0,1]$. Define $\otimes$ and $\rightarrow$ are two partial operations:

$$a \otimes b = \begin{cases} a \wedge b, & if\ a + b \le \alpha\ or\ a = 1\ or\ b = 1 \\ undefined, & others \end{cases}$$

$$a \to b = \begin{cases} 1, & a \le b \\ (\alpha - a) \vee b, & a > b \end{cases}$$

$(a, b \in L)$, where $\alpha \in [0,1]$. Then $(L; \le, \otimes, \to, 0, 1)$ is a PRL, the proper filters are: $\{1\}$ (By Example 5.6, it is similarly that $[x, 1]$, $[1 - x, 1]$ are not filters in $L$, for any $x \in [0,1]$).

**Note:** "$-$" in Definition 6.6 and Definition 6.7 is a substraction operation in the natural sense, which is different from "$-$" in Definition 2.1.

In the following contents, unless otherwise specified, it means that the contents are valid under the condition of definition.

**Proposition 6.8** If $(L; \le, \otimes, \to, 0, 1)$ is a wPRL and $F$ is a filter. Then,

If $x \in F, y \in L, x \to y \in F$, then $y \in F$.

**Proof.** Through known conditions and applying (W2) and (F3), get $x \otimes (x \to y) \in F$, and by Theorem 4.24 (2), $x \otimes (x \to y) \le y$, so, applying (F2), $y \in F$.

**Definition 6.9** Assume $(L; \le, \otimes, \to, 0, 1)$ is a wPRL and $F$ is a filter in $L$. We call $F$ is strong, if it satisfies,

(s1) If $z \to x, z \to y$ are defined and $x \to y \in F$, then $(z \to x) \to (z \to y) \in F$;

(s2) If $y \to z, x \to z$ are defined and $x \to y \in F$, then $(y \to z) \to (x \to z) \in F$;

(s3) If $(x \otimes y) \to z$ are defined and $x \to (y \to z) \in F$, then $(x \otimes y) \to z \in F$.

(s4) If $x \otimes z$, $y \otimes z$ are defined and $x \to y \in F$, then $(x \otimes z) \to (y \otimes z) \in F$.

**Example 6.10** If $L$ = {0,1,2,3}, the induced order ≤ is showed in Fig. 4. The operations ⊗, → are defined by Table 23 and Table 24. Then $L$ is a wPRL. The filters are: {3}, {1,3}, {2,3}; they are not strong filters (Because if $F$ = {3}, it does not meet (s2) and (s4). Then, obviously {1,3} and {2,3} are not strong filters either).

Table 23. The operation ⊗ on $L$

| ⊗ | 0 | 1 | 2 | 3 |
|---|---|---|---|---|
| 0 | | | | 0 |
| 1 | | 1 | 0 | 1 |
| 2 | | 0 | | 2 |
| 3 | 0 | 1 | 2 | 3 |

Table 24. The operation → on $L$

| → | 0 | 1 | 2 | 3 |
|---|---|---|---|---|
| 0 | 3 | 3 | | 3 |
| 1 | 2 | 3 | | 3 |
| 2 | 1 | | 3 | 3 |
| 3 | 0 | 1 | 2 | 3 |

**Example 6.11** In Example 4.20, $(L; \leq, \otimes, \rightarrow, 0,1)$ is a wPRL, the filters are: {3}, {1,3}, {2,3}; they are not strong filters.

**Example 6.12** In Example 4.21, $(L; \leq, \otimes, \rightarrow, 0,1)$ is a wPRL, the filters are: {4}, {2,4}, {3,4}, {1,2,4}; among them, {4}, {3,4} are strong filters, {2,4} and {1,2,4} are not strong filters, because they are not satisfied with (s2) and (s4).

**Example 6.13** In Example 4.22, $(L; \leq, \otimes, \rightarrow, 0,1)$ is a wPRL, the filters are: {5}, {1,5}, {2,5}, {4,5}, {1,2,5}, {1,4,5}, {2,4,5}; they are all strong filters.

**Example 6.14** In Example 4.23, $(L; \leq, \otimes, \rightarrow, 0,1)$ is a wPRL, the filters are: {5}, {2,5}; they are all strong filters.

**Proposition 6.15** If $(L; \leq, \otimes, \rightarrow, 0,1)$ is a wPRL and $F$ is a strong filter in $L$, then,

$$\text{If } (x \otimes y) \rightarrow z \in F, \text{ then } x \rightarrow (y \rightarrow z) \in F.$$

**Proof.** Applying (W2), $(x \otimes y) \otimes ((x \otimes y) \rightarrow z)$ is defined, so $((x \otimes y) \rightarrow z) \otimes (x \otimes y) \leq z$, we have $((x \otimes y) \rightarrow z) \otimes x \leq y \rightarrow z$, hence $(x \otimes y) \rightarrow z \leq x \rightarrow (y \rightarrow z)$. Since $(x \otimes y) \rightarrow z \in F$, $x \rightarrow (y \rightarrow z) \in F$.

**Definition 6.16** Assume $(L; \leq, \otimes, \rightarrow, 0,1)$ is a PRL and $F$ is a filter. For all $x, y \in L$, defined a binary relation $\sim_F$:

$$x \sim_F y \text{ iff } x \rightarrow y \in F \text{ and } y \rightarrow x \in F$$

**Theorem 6.17** If $(L; \leq, \otimes, \rightarrow, 0,1)$ is a wPRL, $F$ is a strong filter in $L$ and $\sim_F$ is a binary relation. Then $\sim_F$ is an equivalent relation.

**Proof.** (1) Because $x \rightarrow x = 1 \in F$, we have $x \sim_F x$.

(2) Applying Definition 6.16, $\sim_F$ is symmetric.

(3) Assume $x \sim_F y, y \sim_F z$. For one thing, $x \rightarrow y \in F$, when $x \rightarrow z$ is defined, by Definition 6.9 (s2), $(y \rightarrow z) \rightarrow (x \rightarrow z) \in F$, $y \rightarrow z \in F$, so $x \rightarrow z \in F$. For another, $z \rightarrow y$, $y \rightarrow x$ are defined, $z \rightarrow y \in F$, when $z \rightarrow x$ is defined, similarly, $(y \rightarrow x) \rightarrow (z \rightarrow x) \in F$, so $z \rightarrow x \in F$. Hence, $x \sim_F z$.

**Definition 6.18** Let $(L; \leq, \otimes, \rightarrow, 0,1)$ be a wPRL. Then $\sim$ is a binary relation, which is called a congruence, for all $x, y, x_1, y_1 \in L$, if:

(C1) $\sim$ is an equivalence relation;

(C2) if $x \sim x_1, y \sim y_1$, $x \otimes y$ and $x_1 \otimes y_1$ are defined, then $(x \otimes y) \sim (x_1 \otimes y_1)$;

(C3) if $x \sim x_1, y \sim y_1$, $x \rightarrow y$ and $x_1 \rightarrow y_1$ are defined, then $(x \rightarrow y) \sim (x_1 \rightarrow y_1)$.

**Theorem 6.19** If $(L; \leq, \otimes, \rightarrow, 0,1)$ is a wPRL and $F$ is a strong filter in $L$, then $\sim_F$ is a congruence relation.

**Proof.** Applying Theorem 6.17, $\sim_F$ is an equivalent relation.

If $x \sim_F x_1$, $y \sim_F y$, and $x \rightarrow x_1 \in F$, then by Definition 6.9 (s4), $(x \otimes y) \rightarrow (x_1 \otimes y) \in F$. Similarly, $(x_1 \otimes y) \rightarrow (x \otimes y) \in F$ can be derived. So, $(x \otimes y) \sim_F (x_1 \otimes y)$ . For the same reason, $(x_1 \otimes y) \sim_F (x_1 \otimes y_1)$. In conclusion, $(x \otimes y) \sim_F (x_1 \otimes y_1)$. This means Definition 6.18 (C2) holds.

From $x \sim x$, $y \sim y_1$ and $y \rightarrow y_1 \in F$, applying Definition 6.9 (s1), $(x \rightarrow y) \rightarrow (x \rightarrow y_1) \in F$, similarly, $(x \rightarrow y_1) \rightarrow (x \rightarrow y) \in F$. Hence, $(x \rightarrow y) \sim_F (x \rightarrow y_1)$. Similarly, applying Definition 6.9 (s2), we can get $(x \rightarrow y_1) \sim_F (x_1 \rightarrow y_1)$. Thus, $(x \rightarrow y) \sim_F (x_1 \rightarrow y_1)$. This means Definition 6.18 (C3) holds.

In the following content, $[x]_F$ represents the equivalence class of $x$, which is for the equivalence relation $\sim_F$. For any $x \in L$, and write the set of all equivalence classes as $L/\sim_F$. Well, we get:

**Definition 6.20** Assume $(L; \leq, \otimes, \rightarrow, 0, 1)$ is a PRL. A set $L/\sim_F = (L/\sim_F; \leq, \otimes, \rightarrow, [0]_F, [1]_F)$, where $F$ is a strong filter, $\sim_F$ is a congruence relation on $L$, $\otimes$ and $\rightarrow$ are partial binary operations on $L/\sim_F$ , $\leq$ is the order relation on $L/\sim_F$, is called a quotient set if, for any $x, y \in L$, the following conditions hold:

$$[x]_F \otimes [y]_F := \begin{cases} [x \otimes y]_F, & \forall a \in [x]_F, b \in [y]_F \text{ and } a \otimes b \text{ is defined} \\ [x]_F, & \text{if } [y]_F = [1]_F \\ undefined, & \exists a \in [x]_F, b \in [y]_F \text{ and } a \otimes b \text{ is undefined} \end{cases}$$

$$[x]_F \rightarrow [y]_F := \begin{cases} [x \rightarrow y]_F, & \forall a \in [x]_F, b \in [y]_F \text{ and } a \rightarrow b \text{ is defined} \\ undefined, & \exists a \in [x]_F, b \in [y]_F \text{ and } a \rightarrow b \text{ is undefined} \end{cases}$$

If $x \rightarrow y$ is defined, then $[x]_F \leq [y]_F$ if and only if $[x]_F \rightarrow [y]_F = [1]_F$.

**Theorem 6.21** Let $\sim_F$ be a congruence relation on $L$. Then $L/\sim_F$ is a PRL.

**Proof.** (PRL1) (1) If for any $a \in [x]_F, b \in [y]_F$ and $a \otimes b$ is defined, then $b \otimes a$ also is, hence $[x]_F \otimes [y]_F = [x \otimes y]_F = [y \otimes x]_F = [y]_F \otimes [x]_F$.

(2) If $[y]_F = [1]_F$, then $[x]_F \otimes [y]_F = [x]_F = [1 \otimes x]_F = [1]_F \otimes [x]_F = [y]_F \otimes [x]_F$.

(PRL2) (1) If for any $a \in [x]_F, b \in [y]_F, c \in [z]_F$ and $b \otimes c$, $a \otimes (b \otimes c)$ are defined, then $a \otimes b$, $(a \otimes b) \otimes c$ are defined, hence $[x]_F \otimes ([y]_F \otimes [z]_F) = [x]_F \otimes ([y \otimes z]_F) = [x \otimes (y \otimes z)]_F = [(x \otimes y) \otimes z]_F = ([x \otimes y]_F) \otimes [z]_F = ([x]_F \otimes [y]_F) \otimes [z]_F$.

(2) ① If $[z]_F = [1]_F$ , then $[x]_F \otimes ([y]_F \otimes [z]_F) = [x]_F \otimes [y]_F = [x \otimes y]_F = [x \otimes y]_F \otimes [z]_F = ([x]_F \otimes [y]_F) \otimes [z]_F$.

② If $[x]_F$ or $[y]_F = [1]_F$, similar proof can be obtained.

③ If $[x]_F$ and $[y]_F \neq [1]_F$, then $[y]_F \otimes [z]_F \neq [1]_F$, i.e., when $[x]_F \otimes ([y]_F \otimes [z]_F)$ is defined, $([x]_F \otimes [y]_F) \otimes [z]_F$ must defined.

Hence, $[x]_F \otimes ([y]_F \otimes [z]_F) = ([x]_F \otimes [y]_F) \otimes [z]_F$.

(PRL3) For all $[x]_F \in L/\sim_F$, $[x]_F \otimes [1]_F = [x]_F$.

(PRL4) Now, we'll prove that $(\otimes, \to)$ is a partial adjoint pair on $L/{\sim_F}$.

(PAP1) If $[x]_F \le [y]_F$, then $[x]_F \to [y]_F = [x \to y]_F = [1]_F$.

(1) If for any $a \in [x]_F, b \in [y]_F, c \in [z]_F$ and $a \otimes c$, $b \otimes c$ are defined, then $[x]_F \otimes [z]_F = [x \otimes z]_F$, $[y]_F \otimes [z]_F = [y \otimes z]_F$. And we know $x \to y \in F$, applying Definition 6.9 (s4), $(x \otimes z) \to (y \otimes z)$ is defined and $(x \otimes z) \to (y \otimes z) \in F = [1]_F$, so $[(x \otimes z) \to (y \otimes z)]_F = [1]_F$, i.e., $[x]_F \otimes [z]_F \le [y]_F \otimes [z]_F$.

(2) If $[z]_F = [1]_F$, then $[x]_F \otimes [z]_F = [x]_F \le [y]_F = [y]_F \otimes [z]_F$.

(PAP2) If $[x]_F \le [y]_F$, then $[x]_F \to [y]_F = [x \to y]_F = [1]_F$. On the one hand, If for any $a \in [x]_F, b \in [y]_F, c \in [z]_F$, then $[x]_F \to [z]_F = [x \to z]_F$, $[y]_F \to [z]_F = [y \to z]_F$, and we know $x \to y \in F$, applying Definition 6.9 (s2), $(y \to z) \to (x \to z) \in F = [1]_F$, so $[(y \to z) \to (x \to z)]_F = [1]_F$, i.e., $[y]_F \to [z]_F \le [x]_F \to [z]_F$.

On the other hand, similarly, applying Definition 6.9 (s1), $[z]_F \to [x]_F \le [z]_F \to [y]_F$.

(PAP3) ($\Rightarrow$) ① If for any $a \in [x]_F, b \in [y]_F, c \in [z]_F$, then $[x]_F \otimes [y]_F = [x \otimes y]_F$, $[x]_F \to [z]_F = [x \to z]_F$. $[x]_F \otimes [y]_F \le [z]_F \Leftrightarrow [x \otimes y]_F \le [z]_F \Leftrightarrow [(x \otimes y) \to z]_F = [1]_F$, i.e., $x \otimes y \to z \in F$, applying Proposition 6.14, $y \to (x \to z) \in F = [1]_F$, so, $[y \to (x \to z)]_F = [1]_F \Leftrightarrow [y]_F \to [x \to z]_F = [1]_F \Leftrightarrow [y]_F \le [x]_F \to [z]_F$.

② If $[y]_F = [1]_F$, then $[x]_F \otimes [y]_F = [x]_F$. If for any $a \in [x]_F, c \in [z]_F$, $[x]_F \to [z]_F = [x \to z]_F$, then $[x]_F \otimes [y]_F \le [z]_F \Leftrightarrow [x]_F \le [z]_F \Leftrightarrow [x \to z]_F = [1]_F$, so, $[1]_F \le [x \to z]_F \Leftrightarrow [y]_F \le [x]_F \to [z]_F$.

($\Leftarrow$) By the same token. Vice versa.

In conclusion, $L/{\sim_F}$ is a partial residuated lattice.

**Example 6.22** Let $L/{\sim_F} = \{\{0,1\}, \{2,3\}\}$, where $F = \{2,3\}$, the induced order $\le$ is showed in Fig. 7. The operations $\otimes$ and $\to$ are defined by Table 25 and Table 26. Then $L/{\sim_F}$ is a PRL.

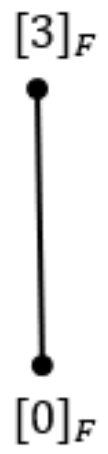

Fig. 7. The order relation on $L/{\sim_F}$

Table 25. The operation $\otimes$

| $\otimes$ | $[0]_F$ | $[3]_F$ |
|---|---|---|
| $[0]_F$ | | $[0]_F$ |
| $[3]_F$ | $[0]_F$ | $[3]_F$ |

Table 26. The operation $\to$

| $\to$ | $[0]_F$ | $[3]_F$ |
|---|---|---|
| $[0]_F$ | $[3]_F$ | |
| $[3]_F$ | | $[3]_F$ |

**Example 6.23** Let $L/{\sim_F}$ ={{0,1}, {2,4}, {3}}, where $F$ = {2,4}, the induced order ≤ is showed in Fig. 8. The operations ⊗ and → are defined by Table 27and Table 28 Then $L/{\sim_F}$ is a PRL.

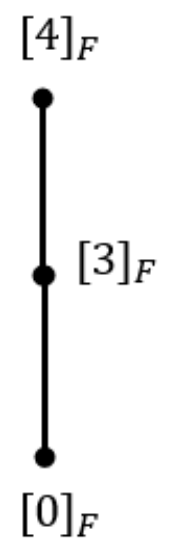

Fig. 8. The order relation on $L/{\sim_F}$

Table 27. The operation ⊗

| ⊗ | $[0]_F$ | $[3]_F$ | $[4]_F$ |
|---|---|---|---|
| $[0]_F$ | | | $[0]_F$ |
| $[3]_F$ | | $[0]_F$ | $[3]_F$ |
| $[4]_F$ | $[0]_F$ | $[3]_F$ | $[4]_F$ |

Table 28. The operation →

| → | $[0]_F$ | $[3]_F$ | $[4]_F$ |
|---|---|---|---|
| $[0]_F$ | | | |
| $[3]_F$ | | $[4]_F$ | |
| $[4]_F$ | $[0]_F$ | | $[4]_F$ |

**Example 6.24** Let $L/{\sim_F}$ ={{0,1}, {2,5}, {3}, {4}}, where $F$= {2,5}, the induced order ≤ is showed in Fig. 9. The operations ⊗ and → are defined by Table 29 and Table 30. Then $L/{\sim_F}$ is a PRL.

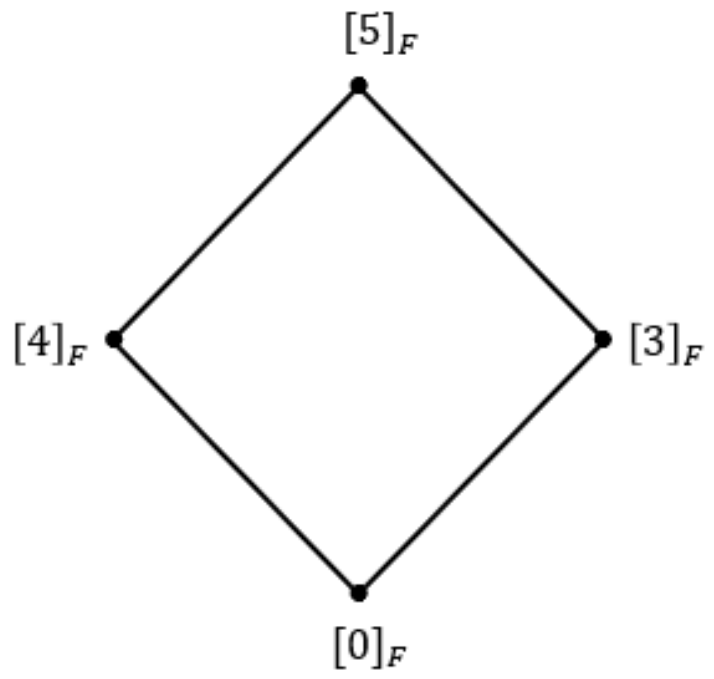

Fig. 9. The order relation on $L/{\sim_F}$

Table 29. The operation ⊗

| $\otimes$ | $[0]_F$ | $[3]_F$ | $[4]_F$ | $[5]_F$ |
|---|---|---|---|---|
| $[0]_F$ | | | | $[0]_F$ |
| $[3]_F$ | | $[0]_F$ | | $[3]_F$ |
| $[4]_F$ | | | $[0]_F$ | $[4]_F$ |
| $[5]_F$ | $[0]_F$ | $[3]_F$ | $[4]_F$ | $[5]_F$ |

Table 30. The operation →

| $\to$ | $[0]_F$ | $[3]_F$ | $[4]_F$ | $[5]_F$ |
|---|---|---|---|---|
| $[0]_F$ | $[5]_F$ | | | |
| $[3]_F$ | | $[5]_F$ | | |
| $[4]_F$ | | | $[5]_F$ | |
| $[5]_F$ | $[0]_F$ | | | $[5]_F$ |

## 7. Conclusions

In this paper, we studied the induced partial residuated implications by partial triangular norms, further explore the relationship between fuzzy logic (PRLs) and quantum logic (lattice effect algebra), and obtain many important conclusions. We have reached some important conclusions:

(1) In lattice effect algebras, define $x \odot y$, we can get $\odot$ is a partial t-norm (see Proposition 2.13).

(2) The operation "$\odot$" in every commutative quasiresiduated lattice is a partial t- norm (see Proposition 2.14).

(3) Conditions for partial residuated implications to be fuzzy implications (see Corollary 3.6).

(4) The methods to constructing PRL from lattice effect algebras (see Theorem 4.15).

(5) $z\mathcal{L}$-PRL (partial residuated lattice in [25]) is a co-residuated lattice (see Theorem 5.19 and Corollary 5.20).

(6) Construction quotient structure $(L/\sim_F; \leq, \otimes, \to, [0]_F, [1]_F)$, and strictly proved it is a partial residuated lattice (see Definition 6.20 and Theorem 6.21).

Moreover, we also put forward some important concepts, such as PRI, PFI, (well, special) PRL, partial residuated co-implication and partial co-residuated lattice.

Finally, we constructed the filter theory of PRL, which can be considered the generalization of the lattice effect algebras filter theory in fuzzy logic and quantum logic. Specifically, we defined filters and congruence relations on partial residuated lattices. And we have established the quotient algebras structure $(L/\sim_F; \leq, \otimes, \to, [0]_F, [1]_F)$ generated by strong filters, which are obtained by adding some specific conditions to the filter.

## Compliance with ethical standards

**Funding:** This study was funded by National Natural Science Foundation of China (Nos. 61976130, 62081240416).

## References

[1] B. D. Baets, Coimplicators, the forgotten connectives, Tatra Mountains Mathematical Publications, 1997, 12: 229-240.

[2] L. Behounek, V. Novák, Towards fuzzy partial logic, 2015 IEEE International Symposium on Multiple-Valued Logic, 2015: 139-144.

[3] L. Běhounek, M. Daňková, Variable-domain fuzzy sets—Part I: Representation, Fuzzy Sets and Systems, 2020, 380: 1-18.

[4] L. Běhounek, M. Daňková, Aggregation Operators with Undefined Inputs or Outputs, International Journal of Uncertainty, Fuzziness and Knowledge-Based Systems, 2022, 30 (01):19-41.

[5] R. A. Borzooei, A. Dvurečenskij, A. H. Sharafi, Material implications in lattice effect algebras, Information Sciences, 2018, 433: 233–240.

[6] P. Burmeister, A Model Theoretic Oriented Approach to Partial Algebras, Berlin, Akademie-Verlag, 1986.

[7] D. Buşneag, P. Dana, Some types of filters in residuated lattices, Soft Computing, 2014, 18 (5): 825-837.

[8] I. Chajda, R. Halas, H. Langer, The logic induced by effect algebras, Soft Computing, 2020, 24 (19): 14275-14286.

[9] I. Chajda, H. Länger, Residuation in lattice effect algebras, Fuzzy Sets and Systems, 2020, 397: 168-178.

[10] A. Dvurečenskij, T. Vetterlein, Congruences and states on pseudoeffect algebras, Foundations of Physics Letters, 2001, 14 (5): 425-446.

[11] X. Fan, X. Zhang, Generalized pseudo-effect algebras and their Riesz congruences and Riesz ideals, Advances in Mathematics, 2010, 39 (4): 419-428.

[12] D. J. Foulis, M. K. Bennett, Effect algebras and unsharp quantum logics, Foundations of Physics, 1994, 24 (10): 1331-1352.

[13] D. J. Foulis, S. Pulmannova, Logical connectives on lattice effect algebras, Studia Logica, 2012, 100 (6): 1291–1315.

[14] B. Jagadeesha, S. P. Kuncham, B. S. Kedukodi, Implications on a lattice, Fuzzy Information and Engineering, 2016, 8 (4): 411-425.

[15] W. Ji, Fuzzy implications in lattice effect algebras, Fuzzy Sets and Systems, 2021, 405: 40–46.

[16] W. Jing, Ideals, filters, and supports in pseudoeffect algebras, International Journal of Theoretical Physics, 2004, 43 (2): 349-358.

[17] S. C. Kleene, Introduction to Metamathematics, New York, D. van Nostrand, 1952.

[18] V. Novak, Fuzzy type theory with partial functions, Iranian Journal of Fuzzy Systems, 2019, 16 (2): 1-16.

[19] Y. Shang, Y. Li, Generalized ideals and supports in pseudo effect algebras, Soft Computing, 2007, 11 (7): 641-645.

[20] T. Vetterlein, BL-algebras and effect algebras, Soft Computing, 2005, 9 (8): 557-564.

[21] Y. Wu, J. Wang, Y. Yang, Lattice-ordered effect algebras and L-algebras, Fuzzy Sets and Systems, 2019, 369: 103-113.

[22] L. A. Zadeh, Fuzzy sets, Information and Control, 1965, 8: 338-353.

[23] X. Zhang, M. Wang, N. Sheng, Q-residuated lattices and lattice pseudoeffect algebras, Soft Computing, 2022, doi:10.1007/s00500-022-06839-w.

[24] M. Zheng, G. Wang, Co-residuated Lattice and Application, Fuzzy Systems and Mathematics, 2005, 19 (4): 1-6.

[25] X. Zhou, Q. Li, Partial residuated structures and quantum structures, Soft Computing, 2008, 12 (12): 1219-1227.

[26] X. Zhou, Q. Li, G. Wang, Residuated lattices and lattice effect algebras, Fuzzy Sets and Systems, 2007, 158 (8): 904-914.

[27] X. Zhang, X. Fan, Pseudo BL-algebras and pseudo effect algebras, Fuzzy Sets and Systems, 2008, 159 (1): 95-106.

[28] X. Zhang, X. Fan, Pseudo-involutive residuated lattices (non-commutative) and pseudo-effect algebras, Fuzzy Systems and Mathematics, 2010, 24 (1): 1–13.